# I-SOAS towards Product Data Management (PDM) based Application's Problems

Zeeshan Ahmed, Detlef Gerhard
*Mechanical Engineering Informatics and Virtual Product Development Division (MIVP)
Vienna University of Technology (TU Wien),
Email: {zeeshan.ahmed, detlef.gerhard} @tuwien.ac.at,
URL: www.mivp.tuwien.ac.at,
Getreidemarkt 9/307 1060, Vienna, Austria*

**Abstract**

*In this research paper we address the importance of Product Data Management (PDM) with respect to the industrial contributional point of view and its major objectives. Moreover we also present some currently available major challenges to the Product Data Management based communities, and targeting those challenges we discuss an already proposed conceptual architectural based helpful approach and briefly describe how this approach can be helpful in solving the PDM communities faced problems.*

## 1. Introduction

Product data management (PDM) is the computer based system which electronically maintains the organizational technical and managerial data to take advantage in maintaining and improving the quality of products and followed development processes [1]. Major objectives of product data management are to improve the quality of products, improve team coordination, deliver products at the time, reduce engineering environment based problems, provide better and secure access to the configuration based information, prevent error creation and propagation by increasing customization of products, efficiently managing the large volumes of engineering data in reusable form generated by computer based systems and reusing design information.

In 1980s, in the begging of PDM proposition, the concept was new and not very welcome by the industry of that time but with the passage of time, PDM is becoming famous and now it is widely in the use of many multinational companies. Now a days, most of the PDM based applications are contributing in industry by providing engineering information management module control to access, store, integrate, secure, recover and manage information in data warehouse, distributed networked computer environment based infrastructure, resource management, information structure management, workflow control and system administration [1]. There are many PDM based application developers like Metaphase (SDRC)[7], SherpaWorks (Inso), Enovia (IBM), CMS (WTC), Windchill (PTC) [8], and Smarteam (Smart Solutions) [6].

In this research paper we are addressing some major available problems to the PDM community in section 2, briefly discussing the current needs of PDM community in section 3, presenting the scope and goals of our research work in section 4, discussing an already intelligent semantic based available and in process conceptual architectural based approach in section 5, justifying the importance of already discussed approach with respect to the need of solutions for currently targeted problems of PDM community in section 6 of the research paper. Then in end concluded and presented some future recommendations in sections 7 of research paper.

## 2. PDM based Application's Problems

Like many other communities of different fields the communities of PDM is also struggling in solving some currently available challenges .i.e., Successful PDM System Implementation, Static and Unfriendly Graphical User Interface, Static and Unintelligent Search, PDM System deployment and Reinstallation, Secure and Scalable, Platform Independent System, Standardized Framework.

*Successful PDM System Implementation;* Successful implementation of a PDM based applications in especially large organizations is a quite difficult, time consuming, expensive and complex task.

It becomes more hectic with respect to the implement point of view when most of the time most of the staff (from corporate management, top level management, engineering management and IT department) do not give importance to the PDM products and without these person's corporation it is quite difficult to gather system requirement based information for PDM based system design implementation [3]. Most of the people (staff) don't want to use PDM based applications because they are afraid for several reasons like top level managers don't want to involve in low level technical and business issues, don't want to spend more money, don't have much time, look for fast pay-back projects and think that the PDM based applications are not matured enough, nonflexible and risky to consider, where as the operational level staff is sometime incapable of handling PDM systems and feel job insecurity.

*Static & Unfriendly Graphical User Interface;* The graphical user interfaces of available PDM based applications are unfriendly and due to this problem it is very hard for a new user to learn and play in short time. Moreover existing PDM based application's GUI's are static, not flexible enough to redesign, slow in event handling and not much intelligent so then it can automatically learn and redesign itself according to the user's ease and need.

*Static and Unintelligent Search;* In every PDM based application, there is a search mechanism required and implemented to locate the user's needed information. But unfortunately still there is no as such intelligent search mechanism available which can process user's dynamic request based queries and can extract the most optimized results in minimum possible time in return.

*PDM System deployment and Reinstallation;* Product Data Management based product are very big and complex, heavily depending upon many third party software for the implementation and due to this reason it's very hard to install full application at once. Moreover PDM Systems are required to be easily extendable because whenever new features are demanded users must most of the time in most of the cases have to reinstall or upgrade the client application completely.

*Secure and Scalable;* Traditional PDM Systems are not adequately available, secure, reliable, and scalable for global enterprise services. Because of highly important technical and managerial data storage and management, it's very important that the PDM system should be highly secured, reliable and scalable so then it cannot easily be hacked and destroyed.

*Platform Independent System;* PDM Systems are required to be platform independent because in the new business model, it is nearly impossible to mandate that all the potential users choose the same platform or the same operation system. Moreover homogeneous network configured PDM needs to provide access to all users placed at different locations, especially those who are on different networks.

*Standardized Framework;* These days many companies are developing many difference PDM systems keeping almost similar goals in their minds but using different framework. It will be a great favor to PDM system development community if there will a standardized framework PDM systems development.

## 3. PDM-Need of the time

Keeping eyes on above discussed major currently faced problem of Product Data Management, we can say, right now PDM community first of all needs a very convincing and strong strategy for clients to win their confidence over the PDM based product development and usage. Moreover PDM community also needs a new approach which can be very helpful in implementing the concepts of Product Data Management in the form of an intelligent knowledge based software application, which must be capable of intelligently handling user's structured and unstructured digital and natural language based requests, process and model the information for fast, optimized and efficient search mechanism, provide options to store, manage and extract heavy information in more optimized and better way, provide a compact installation process to deploy PDM based application in real time environment.

## 4. Scope and Goals of research

Like many other communities of different fields the communities of PDM are also struggling in solving some currently available challenges (mentioned above) but we are addressing only targeted three of all major PDM problems .i.e., Successful PDM System Implementation, Static and Unfriendly Graphical User Interface, Static and Unintelligent Search and Platform Independent System.

Goals of our research work is to support web based platform independent intelligent Product Data Management based Application development which should be capable of Intelligent handling user's requests, Processing and modeling unstructured information and Storing, managing and speedily extracting heavy and light amount of data from repository.

## 5. I-SOAS

To take advantage in solving the problems of implementing intelligent user system communication or intelligent human machine interface development, meta data extraction out of unstructured data, semantic oriented information modeling, fast managed data extraction and final user end data representation an approach Intelligent Semantic Oriented Agent Based Search (I-SOAS) [2] has been proposed. The proposed conceptual architecture of I-SOAS is consists of four main sequential iterative components .i.e., Intelligent User Interface (IUI), Information Processing (IP), Data Management (DM) and Data Representation (DR) [4], as shown in Figure 1.

Intelligent User Interface (IUI) is responsible for the intelligent user system communication. IUI is proposed as an intelligent dynamic user interface capable of first analyzing the source of input, forwarding inputted data for further processing and responding back to the user with end results. Moreover IUI is supposed to be flexible enough so then it can be learned in short time and redesigned by user itself. To implement the IUI as shown in Figure.1, IUI is divided into two main sub-categories .i.e., Graphical User Interface and Communication Sources. Graphical User Interface is consists of the concept of three more sub-categories .i.e., Intelligent, Flexible and Agent to intelligently handle the user's unstructured requests, provide multiple options to redesign the graphical user interface according to the ease of the user by user itself and perform internal architectural component's agent based communication. Where as in Communication Sources, first the corresponding user is supposed to be identified to enable the correct communication mode, if it is a digital system then electronic data communication mode will be enabled and if it is natural system then natural language based communication mode will be enabled [5].

Information Processing (IP) unit is the most important component of the I-SOAS. The quality of performance of I-SOAS depends upon the accuracy in the results of this component. The overall job of IP is divided into five main iterative sequential steps .i.e., Data reading, Tokenization, Parsing, Semantic Modelling, Semantic based query generation. The main concept behind the organization of these five steps is to first understand the semantic hidden in the context of natural or digital set of instructions and generate a semantic information process able model for the system's own understanding and information processing [5]. In the first step, Data Reader is supposed to read and organized inputted data from IUI into initial prioritized instructions list. Then in the second step Data Tokenizer is supposed to tokenized instruction one by one, which are then treated in the third step by Data Parser for parsing and semantic evaluation with respect to the grammar of used natural or digital language. Then in the fourth step Semantic Modeler is supposed to first filter the irrelevant semantic less data and then generate Meta data based semantic model. Then in the last fifth step Semantic Based Query Generator is supposed to generate a new query used further for further data storage and extraction of desired result.

Data Management is responsible for two main functions .i.e.; Semantic based Query Processing and Data Management. Semantic based query built in IP is treated by Semantic based Query Processor to generate

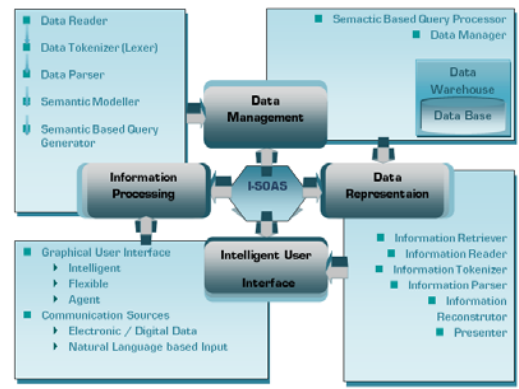

Figure 1. Intelligent Semantic Oriented Agent Based Search (I-SOAS) [4]

SQL query to run in to data base (warehouse) to store and extract the required relevant information. The job of Data Manager is to manage the processes of SQL query building, data extraction and creation of new indexes and storage based on newly retrieved information [5].

Data Representation is responsible for responding back to the user with finalized end results. This component consists of six sub components .i.e., Information Retriever, Information Reader, Information Tokenizer, Information Parser, Information Reconstructor and Presenter. The job of this component is somehow similar to the job of IP, but major difference is of handling data and information. IP treats data to process but DR treats information. Required extracted and managed information from Data Manager is passed to DR using Information Retriever, which simply read and organized by Information Reader without performing any analytical intelligent action except the prioritization of informative statements. Then using Information Tokenizer and Information Parser statements are tokenized and parsed and using Information Reconstructor finalized formatted information is supposed to be build in user's used natural language based format. Finally Presenter presents the resultant information to IUI to respond back to the user [5].

## 6. PDM and I-SAOS

Keeping eyes on above mentioned targeted problems of Product Data Management based applications and solution oriented proposed four components based approach I-SOAS, we can say, that the I-SOAS can be very helpful in solving PDM based problems of Successful PDM Implementation, Static & Unfriendly Graphical User Interface, and Static Search. PDM based applications are looking for an architectural implementable solution for the design implementation of intelligent graphical user interface, which seemed quite similar to the proposition of Intelligent User Interface (IUI) component of the I-SOAS. Moreover PDM based applications are also looking for an implementable solution for the design implementation of an approach which provide the solution to implement dynamic search mechanism to locate user's desired contents in fast and optimized way which is not possible without structuring the unstructured data in a ways that then it can be processed by machine to obtain desired results, which is another goal of I-SOAS to take unstructured data as an input, structure and model it in a way so then it can be processed by the machine. Moreover I-SOAS can also be very helpful in implementing the data management based solution to store, manage and extract information in data base or data warehouse.

## 7. Conclusion

In this research paper we have briefly addressed the importance of Product Data Management (PDM) with respect to the industrial contributional point of view and its major objectives. Moreover we have also presented some currently available major challenges to the Product Data Management based communities, and targeting those challenges we have discussed an already proposed an intelligent semantic based conceptual architectural solution I-SOAS towards PDM communities faced problems. Moreover we have also tried to justify the importance of I-SOAS with respect to the needed solutions for PDM based application's problems.

As this is an in process ongoing research, in future we are aiming to perform in depth analysis of each component of I-SOAS with respect to the architectural and developmental point of view, and will see how practically we can take advantage of I-SOAS in solving the PDM based application's problems.